\documentclass[12pt]{article}
\usepackage{epsfig}
\newcommand{\simlt}  {\raisebox{-.6ex}{$\stackrel{\textstyle <}{\sim}$}}

\def\nl{\hfill\break}
\begin{document}                                                                
\begin{flushright}
RAL-TR/1999-086 \\
hep-ph/9912435 \\
20 December 1999 \\
\end{flushright}                                                               
\begin{center}
{\Large CP and T Violation in Neutrino Oscillations and Invariance of
Jarlskog's Determinant to Matter Effects}
\end{center}
\vspace{2mm}
\begin{center}
{P. F. Harrison\\
Physics Department, Queen Mary and Westfield College\\
Mile End Rd. London E1 4NS. UK \footnotemark[1]}
\end{center}
\begin{center}
{and}
\end{center}                      
\begin{center}                      
{W. G. Scott\\                      
Rutherford Appleton Laboratory\\    
Chilton, Didcot, Oxon OX11 0QX. UK \footnotemark[2]} 
\end{center}
\vspace{2mm}
\begin{abstract}
\baselineskip 0.6cm
\noindent 
Terrestrial matter effects in neutrino propagation are $T$-invariant, so
that any observed $T$ violation when neutrinos pass through the Earth, 
such as an asymmetry between the transition probabilities 
$P(\nu_{\mu}\rightarrow\nu_e)$ and $P(\nu_e\rightarrow\nu_{\mu})$, 
would be a direct indication of $T$ violation at the fundamental level. 
Matter effects {\it do} however modify the magnitudes of $T$-violating
asymmetries, and it has long been known that resonant enhancement can lead
to large effects for a range of plausible values of the relevant
parameters. We note that the determinant of the commutator of the 
lepton mass matrices is invariant under matter effects and use this fact 
to derive a new expression for the $T$-violating asymmetries 
of neutrinos propagating through matter. We give some examples which could 
have physical relevance.
\end{abstract}
\footnotetext[1]{E-mail:p.f.harrison@qmw.ac.uk}
\footnotetext[2]{E-mail:w.g.scott@rl.ac.uk}
\newpage 
\baselineskip 0.6cm
\vskip 1pc
Evidence for oscillations of atmospheric neutrinos \cite{atmos} 
indicates that at least muon and tau neutrinos participate in lepton 
mixing with large amplitudes. Other evidence of a deficit of detected 
solar neutrinos \cite{solar} indicates that electron neutrinos must 
also participate 
in lepton mixing, and the most recent data \cite{SKsolar} disfavour 
small mixing angles. The participation of all three species of neutrinos
in lepton mixing raises the possibility of $CP$ and $T$ violations in
neutrino oscillations, and, given the recent evidence of direct $CP$ 
violation in the hadronic sector \cite{KTEV}, it would perhaps be surprising 
if such effects were not also manifested in the leptonic sector. The 
emergence of large mixing parameters in the leptonic sector offers the 
exciting prospect of 
potentially large $CP$- and $T$-violating asymmetries. One possibility 
\cite{HPS1} which has not been excluded \cite{HPS4} is that 
$CP$ and $T$ violations are maximal for neutrinos in vacuum.
A number of other authors \cite{otherCP} have explored the
phenomenology of $CP$ and $T$ violation in neutrino oscillations for
several different scenarios of lepton mass and mixing parameters.

In both solar and atmospheric experiments, neutrinos can traverse a 
significant fraction of the Earth. As long-baseline accelerator and 
reactor experiments push to still longer baselines, the amount of material 
traversed by man-made neutrino beams also increases. This makes 
the understanding of matter effects in neutrino oscillations essential in 
order to determine the complete pattern of neutrino masses and mixings.
The particle/anti-particle non-invariance of the matter term in the effective 
Hamiltonian for neutrino propagation in matter means that observation of a 
particle-antiparticle asymmetry does not necessarily indicate a fundamental 
$CP$-violation, although it is, in principle, possible to 
discriminate between the two sources of such effects using the observed 
distance- and energy-dependence \cite{otherCPmatter}. However, matter effects 
in neutrino propagation between two points at the surface of the Earth are 
$T$-symmetric, and any observed $T$ violation, such as an asymmetry 
between the transition probabilities $P(\nu_{\mu}\rightarrow\nu_e)$ and  
$P(\nu_e\rightarrow\nu_{\mu})$ would certainly be a direct indication of 
$T$ violation (and via $CPT$ invariance, of $CP$ violation) at the 
fundamental level. 

Despite the fact that matter effects in the Earth are $T$-invariant,
they have a non-trivial effect on the signature of fundamental $T$
violations of $\nu$ and $\overline{\nu}$ separately, 
as well as on $CP$ violations, through their influence on 
the effective neutrino mass and 
mixing parameters. In fact, it has long been known \cite{toshev} that very large
resonant enhancements of the $T$-violating asymmetries are possible in
terrestrial matter.
In this paper, we derive a simple new result for the 
effect of matter on the $T$- and $CP$-violating 
asymmetries, and explore some of the consequences in experimentally 
preferred scenarios of mass and mixing parameters.

In the case of three generations of massive neutrinos, 
in a weak interaction basis which diagonalises the charged lepton mass 
matrix, $M_{\ell} = D_{\ell}$, the neutrino mass matrix, $M_{\nu}$ is in 
general, an arbitrary 3x3 matrix. The Hermitian square of the neutrino 
mass matrix, $M_{\nu}M_{\nu}^{\dag}$,
may be diagonalised to find its eigenvalues, and its eigenvectors form 
the columns of the lepton mixing matrix, U. It is well-known 
that under these circumstances, neutrinos propagating 
in vacuum undergo flavour oscillations, and furthermore, 
in general, these result in $CP$- and $T$-violating asymmetries.

The $CP$- and $T$-violating asymmetries in the transition 
probabilities are given (for arbitrary mixing matrix) by the universal 
function
\begin{eqnarray}
P(\nu_{\alpha} \rightarrow\nu_{\beta}) - P(\overline{\nu}_{\alpha}\rightarrow 
\overline{\nu_{\beta}})
&= &P(\nu_{\alpha}\rightarrow\nu_{\beta}) - P(\nu_{\beta}\rightarrow\nu_{\alpha}) \cr
&= &16J\sin{(\Delta_{12}L/2)}\sin{(\Delta_{23}L/2)}\sin{(\Delta_{31}L/2)}
\label{asymm}
\end{eqnarray}
for any pair of flavour indices $\alpha$, $\beta$, 
where $J$ is Jarlskog's mixing matrix-dependent invariant \cite{cecilia} and the
$\Delta_{ij} = (\lambda_i-\lambda_j)$ are the three differences of 
eigenvalues of the Hamiltonian, $H$. In the vacuum case, it is 
sufficient to set 
\begin{equation}
H = M_{\nu}M_{\nu}^{\dag}/2E
\label{ham}
\end{equation}
so that $\lambda_i = m_i^2/2E$, where $m_i$ is the mass of the $i$th 
neutrino mass eigenstate (as usual, we number the eigenstates 
in increasing order of mass). We note that the parameter $J$ 
has a maximal value of $1/(6\sqrt{3})$, and that the product of the three 
sine functions (the arguments are not independent) has a maximal 
value of $3\sqrt{3}/8$, which it takes when all three arguments are separated 
by $120^{\circ}$ \cite{something}. The maximum magnitude of the product 
of three sines is controlled by the smallest of the three arguments, 
$\Delta_{12}L/2$, and we can consider
$2/\Delta_{12}$ as the reduced wavelength of the asymmetry. The asymmetry is  
observable only once this term has developed a significant phase, and if it 
is furthermore not averaged to zero by resolution effects. This fact limits
the observability of $CP$ and $T$ violations in neutrino oscillations 
in vacuum to a window of parameter-space, as pointed-out in 
Ref.~\cite{schubert}.

In the case that the neutrino beam passes through matter of uniform
density, the Hamiltonian is modified to $H' = H+\Delta H$, where
\begin{equation}
\Delta H = \pm \sqrt{2}GN_e 
\left(\matrix{
1 & 0 & 0 \cr
0 & 0 & 0 \cr
0 & 0 & 0 \cr
} \right).
\label{hamil-b}
\end{equation}
The $\pm$ sign is to be taken as $+$ for neutrino and $-$ for anti-neutrino 
propagation, and makes explicit the particle/anti-particle asymmetry introduced
by matter effects. The effect of matter is therefore to modify the mass 
eigenvalues and the mixing matrix elements, compared with their vacuum values. 

The matter electron density, $N_e$, lies, for the Earth, in the range
$0 \simlt N_e \simlt 6.2 N_A {\rm cm}^{-3}$.
The expression, Eq.~(\ref{asymm}), for the $T$-violating asymmetries is 
still valid in the case of propagation through matter of uniform density, 
except that the eigenvalues and the parameter $J$ appearing in the expression 
are modified to their matter values $\lambda_i'$ and $J'$ respectively. 
For the $CP$-violating asymmetry between neutrinos and anti-neutrinos, 
the situation is made much more complicated by the particle/anti-particle
asymmetry of the matter term in the Hamiltonian, and there are terms 
additional to that on the right-hand side of Eq.~(\ref{asymm}) \cite{otherCPmatter}.

Jarlskog \cite{cecilia} has given an easy way to calculate the $CP$- 
and $T$-violation parameter, $J$, in terms of the mass matrices 
and their eigenvalues. This may be written:
\begin{equation}
2\Delta_{12}\Delta_{23}\Delta_{31}J = {\rm Im}\{{\rm Det}
[M_{\ell}^2,H]\}
/(\Delta_{e\mu}\Delta_{\mu\tau}\Delta_{\tau e})
\label{deter}
\end{equation}
where the $\Delta_{\ell\ell'}$ on the RHS refer to the differences 
between the
squared charged lepton masses. This formula is valid in any weak basis,
but it is useful for our purposes to evaluate the RHS in the weak basis
which diagonalises the charged lepton mass matrix, in which it can be 
written:
\begin{equation}
{\rm Im}\{{\rm Det} [D_{\ell}^2,H]\}
/(\Delta_{e\mu}\Delta_{\mu\tau}\Delta_{\tau e})
= 2{\rm Im}(H_{12}H_{23}H_{31}).
\label{determ}
\end{equation}
It is easy to see that use of the vacuum Hamiltonian, $H$, or the
matter-modified one, $H'$ in Eq.~(\ref{determ}) leaves the
result invariant, as $\Delta H$ is diagonal in this basis. In fact, more
generally, the commutator
\begin{equation}
[D_{\ell}^2,H] = [D_{\ell}^2,H']
\label{inv1}
\end{equation}
is invariant to matter effects in this basis. Taking the determinant on both 
sides of Eq.~(\ref{inv1}) yields the physical result, valid in any weak basis, 
that Jarlskog's determinant of the commutator of the two lepton mass(-squared) 
matrices is invariant to matter effects, and, from Eq.~(\ref{deter}) that:
\begin{equation}
\Delta_{12}\Delta_{23}\Delta_{31}J = 
\Delta'_{12}\Delta'_{23}\Delta'_{31}J'
\label{inv2}
\end{equation}
(where the primed quantities refer to the matter modified Hamiltonian, 
as above). The invariance, Eq.~(\ref{inv2}), is a principal result
of this paper (seeming not to have appeared previously in the literature).
One consequence is that it enables the phenomenology of $T$ violation for 
neutrinos in matter to be stated in a new and transparent way.

The $T$-violating asymmetries of Eq.~(\ref{asymm}) can now be 
generalised to neutrino propagation through matter of uniform density 
in the following form:
\begin{equation}
P'(\nu_{\alpha}\rightarrow\nu_{\beta}) - P'(\nu_{\beta}\rightarrow\nu_{\alpha})
= 16J
\frac{\Delta_{12}\Delta_{23}\Delta_{31}}{\Delta'_{12}\Delta'_{23}\Delta'_{31}}
\sin{(\Delta'_{12}L/2)}\sin{(\Delta'_{23}L/2)}\sin{(\Delta'_{31}L/2)}
\label{newasymm}
\end{equation}
where Eq.~(\ref{inv2}) has been used to rewrite $J'$ in terms of its 
corresponding vacuum value, $J$, the vacuum masses, and the matter-modified
masses, a form which is completely independent of the matter-modified
mixing matrix elements themselves. This result is exact, and valid for
arbitrary vacuum Hamiltonian, ie.~for arbitrary neutrino mass and mixing
parameters. Eq.~(\ref{newasymm}) shows clearly that in general, both the
magnitude, and the wavelength of $T$-violating asymmetries are modified in
matter compared with their vacuum values, and it makes explicit the
correlation between these two effects. The modification of the
magnitude is by a factor:
\begin{equation}
{\cal{R}}=\frac{J'}{J}
=\frac{\Delta_{12}\Delta_{23}\Delta_{31}}
{\Delta'_{12}\Delta'_{23}\Delta'_{31}}, 
\label{enhance}
\end{equation}
which, in general, can be larger or smaller than unity. Provided
that the asymmetry wavelength is not increased by matter effects beyond
observability, potentially useful enhancements of the magnitude of $T$
violation can occur in matter.

In the form of Eq.~(\ref{newasymm}), a number of results concerning 
$T$-violation for neutrinos propagating in matter can be seen rather easily.
For example, it is manifest that even in matter, the $T$-violating asymmetries,
as defined here, are independent of flavour. In the low-density limit, 
$N_e\rightarrow 0$ ($\Delta'_{ij} \rightarrow \Delta_{ij}$ for all $i,j$), 
the expression, Eq.~(\ref{newasymm}), reduces to the normal vacuum case, 
Eq.~(\ref{asymm}), as expected. 
Furthermore, in the small-$L$ limit (ie. if the propagation distance is small 
compared with the scale of the shortest matter oscillation length, 
$L<<2/\Delta'_{13}$), the equation reduces to the vacuum case, 
Eq.~(\ref{asymm}), as all the sine functions in Eq.~(\ref{newasymm}) may
be approximated by their arguments and all the primed variables cancel.
For $L \rightarrow 0$ therefore, there is no observable effect due to matter 
on $T$-asymmetries (nor indeed on any observable aspect of the oscillation
phenomenon \cite{barger}).
In general, if $J \neq 0$ and there are 
no degeneracies in vacuum, Eq.~(\ref{inv2}) 
ensures that there are no degenerate eigenvalues in matter 
($\Delta'_{12}\Delta'_{23}\Delta'_{31} \ne 0$) and hence that ${\cal{R}}$ 
remains finite. 

We note that for an arbitrary Hermitian matrix, $H$, 
the combination of its eigenvalues, $\Delta_{12}\Delta_{23}\Delta_{31}$,
is the square-root of the discriminant of its eigenvalue 
equation:
\begin{equation}
\Delta_{12}\Delta_{23}\Delta_{31} = \sqrt{-4{\cal{S}}^3+{\cal{S}}^2
{\cal{T}}^2-27{\cal{D}}^2+18{\cal{D}}{\cal{T}}{\cal{S}}-4{\cal{D}}{\cal{T}}^3}
\label{discrim}
\end{equation}
where the invariants $\cal{T}$, $\cal{S}$ and $\cal{D}$ are respectively 
the trace, the sum of the principle minors and the 
determinant of $H$. This can be further simplified by noting that we
have the freedom to add an arbitrary multiple of the unit matrix to
$H$ without altering its discriminant, so that we can choose either
${\cal{T}}=0$ or one of the eigenvalues, $\lambda_i=0$.
So, for any given model of the neutrino masses and mixing angles in
vacuum, the magnitudes of the $T$-violating asymmetries in matter can,
in principle, be calculated directly from the vacuum parameters of the model 
and the (simplified) invariants of the matter-modified Hamiltonian,
$H'$, thereby obviating the need to diagonalise explicitly $H'$
to find the matter-modified mixing angles.

In the general case, application of Eq.~(\ref{discrim}) to the matter-modified 
Hamiltonian to calculate the denominator of Eq.~(\ref{enhance}) yields 
the square-root of a quartic function of $N_e$, with coefficients which are 
functions of the vacuum mixing parameters (masses and mixing angles) and the 
neutrino energy. The presence of this quartic function means that 
${\cal{R}}$ has either one or two resonant maxima, as a function of the 
matter density. These correspond to the well-known matter resonances, 
occuring approximately where pairs of matter-modified neutrino masses are 
most nearly degenerate \cite{barger}. Values of ${\cal{R}}$ can be arbitrarily 
large, if $J$ is sufficienctly small (clearly, $J'=RJ$ cannot exceed 
$1/6\sqrt{3}$). For very large values of $N_e$, the quartic term dominates, 
and $T$ asymmetries are suppressed by $\sim 1/N_e^2$.

In the remainder of this paper, we explore some specific examples
which could have physical relevance.
In order to ensure that our discussion is relevant to experiment, 
we will restrict our considerations to regions of parameter
space which are not excluded by present, corroborated neutrino
experiments, such as the Super Kamiokande atmospheric neutrino data \cite{atmos} and the 
Super Kamiokande and Gallium solar neutrino 
data \cite{solar}. We will use for the lepton mixing matrix, $U$, 
a conventional form in which there are three real mixing angles 
and one complex phase such that: 
$U_{e3}=\sin{\theta_{13}}$, $U_{e2}=\sin{\theta_{12}}\cos{\theta_{13}}$, 
$U_{\mu 3}=\sin{\theta_{23}}\cos{\theta_{13}}e^{i\delta}$ and all the
other elements are fixed by unitarity.

A priori, our preferred scenario for neutrino oscillations is the threefold 
maximal, or tri-maximal scheme \cite{HPS1}, which still provides a broadly 
consistent account of all the
corroborated sets of experimental data on neutrino oscillations.
In this scheme, all the elements of the lepton mixing matrix have equal
moduli of magnitude $1/\sqrt{3}$, the vacuum value of $J$ takes its
maximal value, $1/(6\sqrt{3})$, and all observed neutrino disappearance 
data are the result of neutrino oscillations governed by the scale of the 
larger neutrino mass-squared difference, $\delta m_{13}^2 \sim 10^{-3}$ eV$^2$,
with $\delta m_{12}^2$ unresolved even by the 
solar data. This scheme is summarised by 
$\theta_{12}=\theta_{23}=45^{\circ}$, $\theta_{13}=\sin^{-1}{(1/\sqrt{3})}$
and $\delta = 90^{\circ}$. In this case, ${\cal R}$ is always less than 
unity in matter, so that asymmetries are always suppressed. 

A viable, and even experimentally favoured, alternative to tri-maximal mixing 
effectively factorises the atmospheric and solar scales by setting 
$U_{e3} \simeq 0$ in the mixing matrix \cite{bimax}.
The atmospheric data then require $\theta_{23} \simeq 45^{\circ}$, and
an energy-independent solar supression of $1/2$ is obtained by 
setting $\theta_{12} = 45^{\circ}$ in the original bi-maximal 
scheme \cite{bimax}. We have ourselves proposed \cite{HPS4} 
(see also \cite{billsolo}) a particular variant of the bi-maximal scheme with 
$U_{e2}=U_{\mu2}=U_{\tau2}=1/3$ which simulates very effectively 
tri-maximal mixing with a solar supression of 5/9.
In bi-maximal-type schemes, $J_{CP}=0$ and there are no fundamental $CP$- 
or $T$-violating asymmetries involving neutrinos, in vacuum or in
matter. There are still fake $\nu/\bar{\nu}$ asymmetries due to 
matter effects, but these are of little fundamental interest.

For illustrative purposes, we have explored in this paper an ansatz which 
interpolates smoothly between tri-maximal and bi-maximal mixing:
\begin{equation}
U = \left(\matrix{                                                                
\frac{1}{\sqrt{2}}\cos{\theta_{13}} & \frac{1}{\sqrt{2}}\cos{\theta_{13}} & \sin{\theta_{13}} \cr
-\frac{1}{2}(1+\sin{\theta_{13}}e^{i\delta}) & \frac{1}{2}(1-\sin{\theta_{13}}e^{i\delta}) & \frac{1}{\sqrt{2}}\cos{\theta_{13}}e^{i\delta} \cr                                        
\frac{1}{2}(1-\sin{\theta_{13}}e^{i\delta}) & -\frac{1}{2}(1+\sin{\theta_{13}}e^{i\delta}) & \frac{1}{\sqrt{2}}\cos{\theta_{13}}e^{i\delta} \cr
}\right)
\label{mixmat2}
\end{equation}
with $\delta = 90^{\circ}$.
We have assumed that $\delta m_{13}^2 \sim 10^{-3}$ eV$^2$ and have allowed  
$\delta m_{12}^2$ to remain variable. This scheme has the following 
properties:
\begin{itemize}
\item{$\nu_{\mu}$ and $\nu_{\tau}$ are treated democratically}
\item{$\nu_1$ and $\nu_2$ are treated democratically}
\item{in the limit $\sin^2{\theta_{13}} \rightarrow 1/3$, we obtain 
tri-maximal mixing}
\item{in the limit $\sin^2{\theta_{13}} \rightarrow 0$, we obtain 
bi-maximal mixing}
\item{$J_{CP} = (1/4)\sin{\theta_{13}}\cos^2{\theta_{13}}$ varies between
its minimal and maximal values, as we move from bi-maximal to tri-maximal 
mixing.}
\end{itemize}
For arbitrary $\sin{\theta_{13}} (0 < \sin{\theta_{13}} < \frac{1}{\sqrt{3}})$, 
there is always a resonance where $T$-violating asymmetries are maximised
and this can be probed by choosing the neutrino energy, $E$, 
or the density of matter traversed, appropriately, as long as 
there is sufficient pathlength for the asymmetry to develop. The values
of neutrino energy and/or matter density at resonance, and the maximum 
magnitude of the asymmetry there depend on $\sin{\theta_{13}}$ and the 
vacuum masses.

Figs.~1a and 1b show some examples of the maximum magnitude of the 
$T$-violating asymmetry defined in Eq.~(\ref{newasymm}), as a function of the 
matter density in units of $N_A$ cm$^{-3}$ for several values of 
$\sin{\theta_{13}}$ in this ansatz. The two figures differ in terms of
the hierarchy of vacuum mass values used, and illustrate cases with one 
and with two resonant densities respectively. All the curves are for a fixed 
neutrino energy, $E=1.5$ GeV, which turns out to be the most suitable energy 
scale to maximise these effects in terrestrial matter within this model 
(see Eq.~(\ref{result2})). The Earth's mantle has an electron density of 
approximately $2 N_A$ cm$^{-3}$ and the core, roughly $6 N_A$ cm$^{-3}$. The 
negative density parts of the curves correspond to anti-neutrinos propagating 
in normal matter with electron density $|N_e|$. We note that enhancements for
neutrinos, for example, are typically compensated by suppressions for
anti-neutrinos, and/or for neutrinos at different energies.
As shown in Ref.~\cite{HPS4}, the 
electron neutrino begins to decouple completely in the high energy limit, 
and $T$ asymmetries are suppressed accordingly. The magnitudes of the 
asymmetries in vacuum can be read from the curves at the zero density point.  
Several other features are typical of the generic MSW-like matter effect, 
eg., the resonance gets sharper and the asymmetry 
maximum gets closer to unity as the vacuum mixing angle decreases.

It is interesting to consider under what circumstances the
asymmetry becomes maximal in this ansatz. Even in the limit 
$\sin{\theta_{13}}\rightarrow 0$ ($\sin{\theta_{13}} \neq 0$), 
where $T$- and $CP$-violating asymmetries in vacuum become arbitrarily small, 
they can reach their maximum possible value of unity in matter.
We find that this can be achieved if
\begin{equation}
\delta m_{12}^2=2\cos{\theta_{13}}\frac{\sqrt{2}-\cos{\theta_{13}}}{1+\sin^2{\theta_{13}}}\delta m_{13}^2
\label{result1}
\end{equation}
and then the matter density and neutrino energy at resonance are related by:
\begin{equation}
\sqrt{2}GN_e =\frac{1}{\sqrt{2}}\frac{(1-3\sin^2{\theta_{13}})(\sqrt{2}-\cos{\theta_{13}})}{1+\sin^2{\theta_{13}}}\delta m_{13}^2/2E.
\label{result2}
\end{equation}
Under the above two conditions, threefold maximal mixing is achieved at
resonance. Eq.~(\ref{result1}) implies that for such mixing at resonance, the
two $\Delta m^2$ values must be of the same order of magnitude, which is
probably not ruled out by the present data.

Fig.~2 shows an example of how one of the asymmetries in 
Fig.~1 develops with propagation distance, in units of 1000 km. It is
plotted for neutrinos in matter with a constant density of $1.9 N_A$ cm$^{-3}$,
ie. close to the density of the Earth's mantle. The 
matter-enhanced asymmetry is compared with the vacuum asymmetry and the 
corresponding matter-enhanced asymmetry for anti-neutrinos. The enhanced 
asymmetry does not develop to the 100\% level within the distance scale 
of the Earth's diameter in this case, (it would, given a longer pathlength), 
although there are cases where it does. It does however exceed
considerably the vacuum asymmetry, along almost the whole trajectory within 
the Earth, and may be observable when the vacuum asymmetry would not be.

While Eq.~(\ref{result1}) may not be satisfied exactly in nature, there is
a significant range of parameter space over which large enhancements are
possible. We have found the conditions under which the enhancements are
maximal, and here, at least for some values of the parameters, matter can 
induce a situation where the mixing matrix is arbitrarily close to the 
tri-maximal form. Such a scenario might well be obtained in nature, and we 
recommend that large $T$- and $CP$-violating asymmetries be searched for in 
future long-baseline neutrino experiments. It may even be possible to site
experiments to exploit the matter effects in the Earth so as to maximise 
the asymmetries.

\newpage
\noindent {\bf {\large Figure Captions}}

\vspace{10mm}
\noindent Figure~1.
Examples of resonant enhancement and suppression of $T$ violation for neutrino
oscillations in matter using the ansatz of Eq.~(\ref{mixmat2}). The maximum 
magnitude of the $T$-violating asymmetry is plotted as a function of matter 
density and $\sin{\theta_{13}}$, for neutrino energy, $E = 1.5$ GeV, with 
a). $\delta m_{13}^2=1.0\times 10^{-3}$ eV$^2$ and 
$\delta m_{12}^2=0.7\times 10^{-3}$ eV$^2$;
b). $\delta m_{13}^2=1.0\times 10^{-3}$ eV$^2$ and 
$\delta m_{12}^2=0.2\times 10^{-3}$ eV$^2$.
In each case, the point marked ``o'' represents the vacuum value of the 
asymmetry for the same vacuum mass parameters. The Earth's mantle has an 
electron density of approximately $2 N_A$ cm$^{-3}$ and the core, roughly 
$6 N_A$ cm$^{-3}$. NB. The density scale can be converted to
a scale of energy in GeV for neutrinos propagating in the Earth's mantle by 
multiplying the numbers by $\sim 0.75$.

\vspace{10mm}
\noindent Figure~2.
An example of resonant enhancement of $T$ violation for neutrino
  oscillations in matter. The solid curve shows the 
  $T$-violating asymmetry as a function of propagation distance, for neutrinos
  of energy, $E=1.5$ GeV, with $\delta m_{13}^2=1.0 \times 10^{-3}$ eV$^2$,  
  $\delta m_{12}^2=0.7\times 10^{-3}$ eV$^2$ and $\sin{\theta_{13}=0.1}$.
  The matter electron density in this example is $1.9 N_A$ cm$^{-3}$. 
  The dashed line shows the same quantity in vacuum for the same vacuum 
  parameters, and the dotted line shows the same quantity for
  anti-neutrinos.

\newpage
\begin{figure*}
  \begin{center}
     \mbox{\epsfig{file=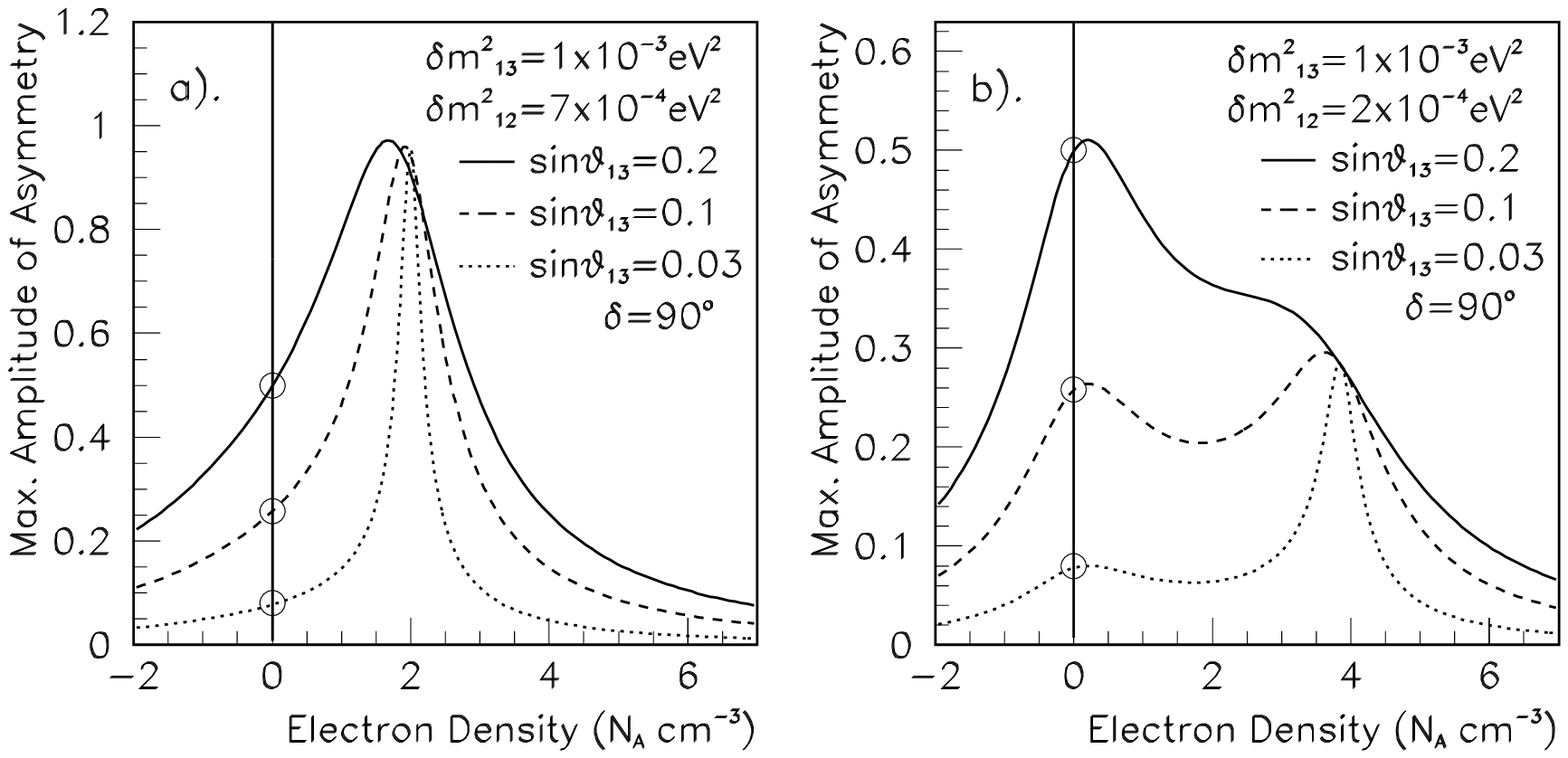, width=24cm, angle=270}}
  \end{center}
\end{figure*}
\newpage
\begin{figure*}
  \begin{center}
     \mbox{\epsfig{file=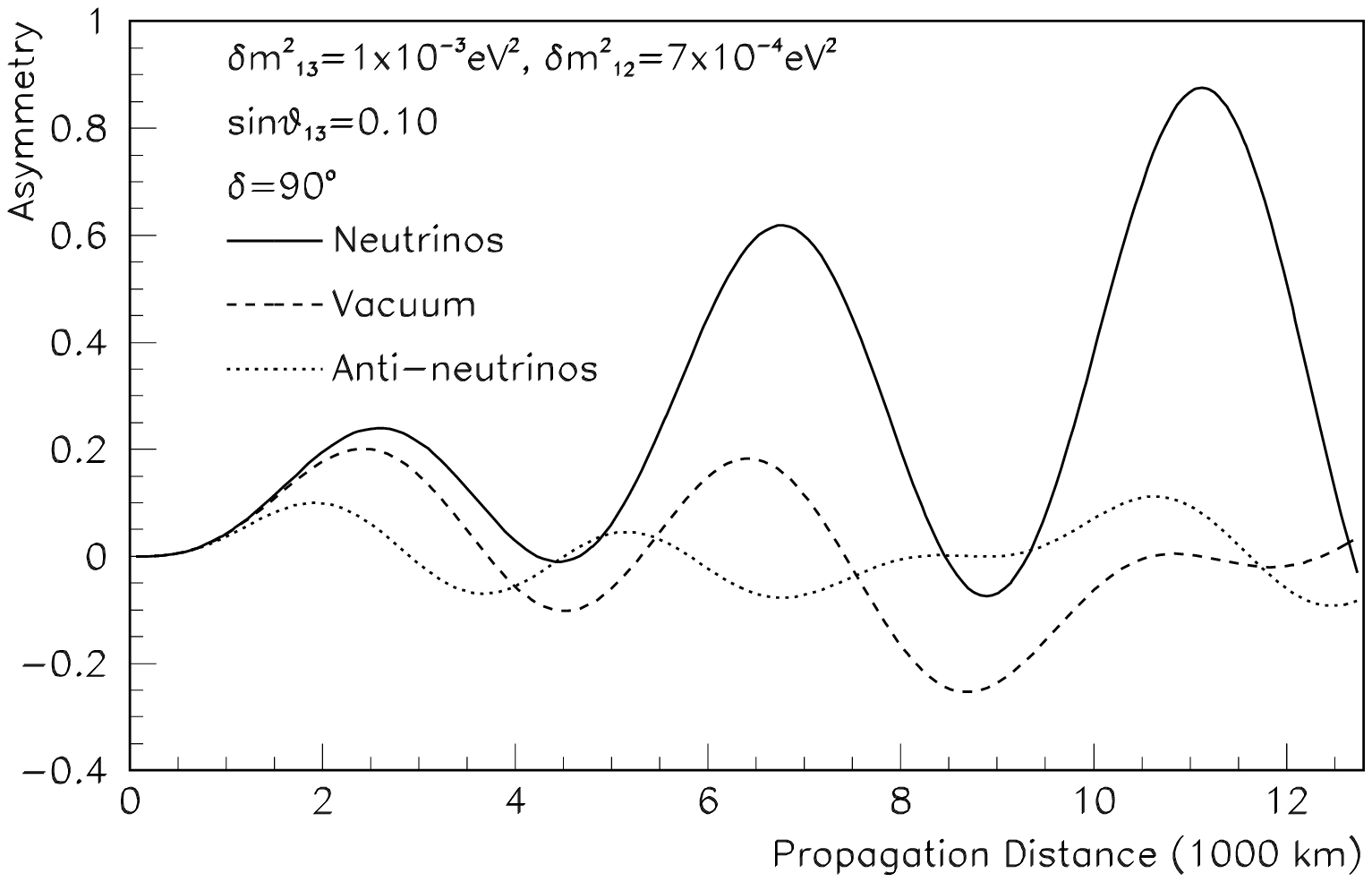, width=22cm, angle=270}}
  \end{center}
\end{figure*}

\end{document}